\documentclass[11pt,a4paper]{article}
\usepackage{jcappub}

\usepackage{esint}

\title{Initial Conditions in the Effective Field Theory of Inflation}
\author{Ross O'Connell,}
\author{R. Holman}
\affiliation{Physics Department, \\
Carnegie Mellon University\\
Pittsburgh PA 15213 USA}

\emailAdd{rcoconne@andrew.cmu.edu}
\emailAdd{rh4a@andrew.cmu.edu}

\abstract{
Many different models of inflation give rise to the same effective field theory of the inflaton. While effective field theories in flat space provide little information about UV physics, we propose that in inflationary backgrounds a large amount of information can be encoded by the initial conditions of the effective theory.  We identify conditions under which this information remains available at late times, e.g. through observation of non-Gaussianities, and introduce a simple model of initial conditions where these conditions are satisfied.
}

\arxivnumber{1109.1562}

\begin{document}
\maketitle

\section{Introduction}


The techniques of effective field theory (EFT) \citep{Polchinski1992,Georgi1993,Rothstein2003,
Burgess2007} allow us to separate our description of physics into a low-energy part, i.e. that below an energy threshold or cutoff $M$, and a high-energy piece containing physics above $M$. An important aspect of this procedure is that high-energy effects only show up in the low-energy description in the form of {\em irrelevant} operators containing only low-energy fields and suppressed by the scale $M$. 
In the diagrammatic approach, these operators are generated by integrating out the high-energy fields when they mediate interactions between low-energy fields.

There are then two possible ways to make use of EFT techniques. When we already know the {\em full} theory we can attempt to determine the exact form of the irrelevant operators in the low-energy theory via an explicit 
calculation in the full theory. However, the situation in particle physics is that we {\em do not} have the full theory, and can at best constrain its form by looking at the effects of irrelevant operators consistent with the known symmetries of the standard model, with arbitrary coefficients, on low energy observables. Needless to say, it is quite difficult to arrive at any definite conclusions about possible UV completions of the standard model in this way.

A similar situation obtains in cosmology. There, we have evidence that something akin to an inflationary era happened in the early universe,  with the concomitant generation of metric perturbations from quantum fluctuations of the inflaton field \cite{Larson2011} . An effective theory approach to inflaton fluctuations is possible \cite{Cheung2008} and, like the EFT's that try to describe beyond-the-standard model physics, the effective theory for inflation also has a cutoff $M$
 associated with it. However, the expansion of the Universe brings in new features that we wish to explore in this work.

To what extent could we access information above the scale $M$?
If we were in flat space, our only options would be to either start doing experiments at energies at or above the cutoff or, if that is not possible, to try to pin down some of the coefficients of the irrelevant operators induced by the UV completion using low-energy experiments. Until the advent of the LHC the first approach was not possible, while the second is a thoroughly arduous task, as witnessed by the fact that roughly 20 years of this sort of work on the electroweak theory yielded virtually no information about the mechanism of electroweak symmetry breaking.

In inflation, we could try to use low-energy data to determine the coefficients of the irrelevant operators. However, we are stymied to a great extent in this task by the fact that there are not that many low-energy observables that are directly relevant to inflaton fluctuations. In fact, we are restricted mostly to data coming from the CMB. We can study the power spectrum, related to the two point function of the inflaton fluctuations, or use measures of non-Gaussianity, such as the three and higher point functions to access information about the inflaton, but it is not hard to see that this approach can provide only a small amount of information about UV physics. Furthermore, if the scale of inflation is of order $\sim 10^{14}\ {\rm GeV}$, any attempt to study the theory directly above the cutoff scale would be futile.

In this work we identify an approach to could in principle yield detailed information about physics above the cutoff of the EFT.
The crucial observation is that the phenomenon of inflation involves
both the dynamics of the inflaton \emph{and} a set of initial conditions \citep{Martin2001,Danielsson2002,Burgess2003,Greene2005,Easther2005}. In the flat space case the initial state is always chosen to be the (either in or out) vacuum or particle states built from it, and there is no ``flow'' from high-momentum to low-momentum. Cosmology in an expanding universe, on the other hand, allows for the redshifting of momenta above the cutoff to momenta below it. Coupling this to the  EFT point of view, which forces us to only consider degrees of freedom below the cutoff, we see that specifying the initial conditions when a mode has physical momentum above $M$ makes no sense.  In particular, if we were to specify initial conditions on a spacelike hypersurface that 
included $t\to-\infty$, some modes would necessarily have physical momenta greater than
the cutoff. Thus the EFT requires us to set initial conditions when the modes cross into its domain of validity, i.e. when $k=a(t_k) M$. These initial conditions naturally encode some information about physics above the cutoff. We will argue that under a relatively mild set of assumptions, information about the initial conditions for a range of modes remains available at late times. The $k$-dependence of the initial conditions then provides much more detailed information about UV physics than we would usually expect access to from an effective field theory.

In section \ref{EFT} we review the basic properties of EFTs, with special attention to the problem of discriminating between possible UV completions.  In section \ref{ICs} we  examine when non-Bunch-Davies initial conditions remain observable at late times.  We find that small deviations from Bunch-Davies can be obscured by quantum noise, while large deviations can thermalize. However, there is a potential ``sweet spot'' in which information from the inflaton initial state can be potentially observable, although it is difficult to be more specific than that without further information.  We introduce a simple model for the initial conditions, where the ``sweet spot'' is realized, explicitly demonstrating that inflation does \emph{not} necessarily eliminate all information about the initial conditions of the inflaton.


In section \ref{EX} we consider the prospects for observing non-Bunch-Davies initial conditions.  A basic challenge is that the space of possible initial conditions does not have any \emph{a priori} constraints, so we focus on the (essentially universal) effects of thermal freeze-out of the inflaton. We use the WMAP7 power spectrum \cite{Larson2011} to constrain this situation, and find that it is largely (but not completely) excluded. Specifically, we find that when the span between the Hubble scale and the EFT cutoff is set by $H/M=10^{-4}$, the largest momentum affected by thermal freeze-out must be less than $10^{-3}\,\mathrm{Mpc}^{-1}$, while the smallest momentum included in the CMB sky is roughly $10^{-4}\,\mathrm{Mpc}^{-1}$.

\section{Effective Field Theory and Cosmology}
\label{EFT}

Because the effective field theory point of view is central to our argument, we begin by reviewing the properties of effective field theories in flat space.  We then describe how this picture changes in inflationary backgrounds.  While it is widely appreciated that the EFT for inflaton fluctuations in \cite{Cheung2008} (henceforth CCFKS) provides a unified description of inflationary models, the existence of an EFT also has important conceptual implications for the study of inflation.  Among these is that so long as the \emph{dynamics} of the EFT of inflation are consistent with observations, more careful study of the dynamics provides virtually no information about the UV completion of the theory\footnote{The UV physics we endeavor to study is not necessarily trans-Planckian, since the cutoff of the EFT is not necessarily the Planck scale.}.  This is our principal motivation for studying the \emph{state} of the theory.

\subsection{The Structure of Effective Field Theories}

We begin by considering a weakly-coupled scalar field in flat space.  Generically, the Lagrangian takes the form 

\begin{equation}
\label{eq:philagrangian}
  \mathcal{L}= \left(\partial \phi \right)^2 + \sum_i \frac{g_i}{M^{d_i -4}} \mathcal{O}_i^{(d_i)} \, .
\end{equation}
We assume that a symmetry (e.g. $\phi \to \phi+b$) forbids a mass term $m^2 \phi^2$.  In addition to the kinetic term, there is an infinite tower of interactions encoded by the $\mathcal{O}_i^{(d_i)}$, with dimension $d_i \ge 4$.  At energies $E\ll M$, the series converges quickly, and in practice we can neglect all but a handful of terms. At $E\sim M$ the series breaks down and the EFT requires some sort of UV completion and, as discussed in the introduction, there are two basic strategies for learning about the UV completion.  The first is to perform experiments in the regime where the EFT breaks down.  If this is impractical, one can also do more detailed studies at energies below $M$, determining a few of the $\mathcal{O}_i^{(d_i)}$ and their coefficients $g_i$. 

The organization of the Lagrangian described above also applies in inflationary backgrounds. Although we focus on a different set of observables, the EFT description of the dynamics still breaks down near the cutoff.  In flat space we focus on the S-matrix, which typically becomes non-unitary when the momentum exchanged approaches $M$. In inflationary backgrounds we typically use the interaction Hamiltonian to evolve states in time, but the perturbative expression for the interaction Hamiltonian breaks down when the physical momentum of the state approaches $M$ -- in other words, a state with finite energy at finite time cannot be evolved into the infinite past.  

At first glance, it would appear that we have significantly less access to UV physics when studying inflation.  While we have steadily improved our understanding of particle physics by increasing the energies accessible at colliders, the primary tool we have for studying inflation is the CMB, which only provides information about modes when they have physical momentum $H$. Since we can only observe at one scale, it would seem that the only information available about UV physics comes through the $\mathcal{O}_i^{(d_i)}$. If this were true, we would only be able to make very gross distinctions between UV models. On the other hand, each mode that we observe in the CMB had, at one point, a physical momentum close to $M$.  If some sort of signal were impressed on the mode when its physical momentum was near the cutoff, and that signal survived until the mode crossed the Hubble horizon, information about UV physics would be much \emph{more} accessible than it is in flat space.  What we have described is a question about initial conditions, reformulated to be consistent with the principles of effective field theory.

\subsection{The Effective Field Theory of Inflation}

The EFT for inflaton fluctuations \cite{Cheung2008} 
views the inflation fluctuations $\pi$ as the Goldstone boson of the time reparametrization symmetry, which is broken by the choice of an inflaton zero mode. From this perspective, the fluctuations behave  as the longitudinal gauge mode in spontaneously broken gauge theories. There, at sufficiently high energies, the dynamics of the longitudinal gauge degree of freedom can be described by that of the would-be Goldstone boson. In the inflationary case, the analogous situation is that in the high energy limit, namely when $E\gg E_{\mathrm{mix}}=\sqrt{\epsilon}H$, mixing with the tensor modes can be neglected, as can all the non-derivative suppressed interactions of $\pi$.

The logic is to first choose a gauge for which $\pi=0$, and construct the most general form of the action which is invariant under {\em time-dependent spatial} diffeomorphisms, organized as fluctuations about a background solution. Then, by reintroducing the time reparametrization gauge transformation a la Stueckelberg, the action can be made fully diffeomorphism invariant, albeit in a non-manifest way.

In the high energy limit the fluctuations are governed by the following effective action up to cubic order
\begin{equation}
\label{eq:action1}
S=\int d^4 x \sqrt{-g} \left[ \left(\frac{1}{c_s^2}\dot{\pi}^2 -\frac{(\partial_i \pi)^2 }{a^2}\right) +\frac{\dot{\pi}^3}{M^2}+\cdots   \right].
\end{equation}
Here $c_s$ is the effective speed of sound for the fluctuations (taken to be near unity), $\pi$ is the canonically normalized Goldstone field and we have used the quadratic equation of motion to simplify the cubic term. This action gives a general effective field theory for inflation, from which one can recover specific models by a judicious choice of the parameters. 

The restriction to high energies means that the decoupled effective theory will break down shortly after horizon crossing. However, the simple relation $\pi = -\zeta\slash H$ between $\pi$ and the curvature perturbation $\zeta$, valid in single field inflation, lets us study the complete history of a mode. We first use the CCFKS action to follow a given $\pi$ mode up to and just beyond horizon crossing, and then use the constancy of $\zeta$ outside the horizon.

\section{Initial Conditions}
\label{ICs}

Having argued that non-standard initial conditions could provide a unique window on UV physics, we now examine a set of necessary conditions for the initial conditions to be observable at late times (e.g. horizon crossing).  Generic initial conditions for the inflaton fluctuations can be written as a Bogoliubov
transformation of the Bunch-Davies vacuum, i.e. we expand the field
in modes $v_{k}\left(a\right)$ that are written in terms of the Bunch-Davies
modes $u_k\left(a\right)$ as 
\[
v_{k}\left(a\right)  =  \alpha_{k}u_{k}\left(a\right)+\beta_{k}u_{k}^{*}\left(a\right)\,,
\]
with the coefficients satisfying 
\[
\left|\alpha_{k}\right|^{2}-\left|\beta_{k}\right|^{2}=1\,.
\]
The number of excitations of momentum $k$, relative to the Bunch-Davies vacuum,
is then given by $\left|\beta_{k}\right|^{2}$. Different
models of the initial conditions\footnote{In a free theory, the choice $\beta_k=\sinh \alpha$ reproduces the "alpha states" of \cite{Einhorn2003,Einhorn2003a,Collins2003,Collins2004,Collins2005}, a quantum consistent version of the alpha vacua \cite{Mottola:1984ar,Allen:1985ux}.  In an interacting theory imposing the same conditions at different times leads to different states.} then correspond to different choices
of the $\beta_{k}$ at $a=M/k$.

While initial conditions are traditionally imposed on a spacelike hypersurface at $t\to -\infty$, this is not possible in the EFT of inflation. As $t\to -\infty$ the physical momenta of all modes diverge, and are thus outside of the range of validity of the EFT. Without a usable expression for the interaction Hamiltonian, we cannot evolve the state forward in time.  While one might be able to evolve the state forward with a candidate UV completion of the EFT, we do not believe that currently available data allows one to credibly claim that a particular UV completion is in any way favored over any others.  Instead, it is appropriate to specify the state of a mode (i.e. the $\beta_l$) as it crosses into the EFT.  This proposal has been made by other authors, for example \cite{Danielsson2002,Easther2002}. 

Two important physical effects can obscure the initial conditions.
First, when the density of excitations is sufficiently high they can
thermalize. For modes that reach equilibrium, information about the
initial conditions is lost, and the $\beta_{k}$ are given by a thermal
distribution. In order to determine which modes thermalize, we begin
from Eq.\eqref{eq:action1}. Despite the expansion of the background cosmology, we can compute a scattering cross section, as long as the mean free path is shorter than the Hubble length $H^{-1}$. Doing this, we find that parametrically the cross-section
for $\pi-\pi$ scattering is 
\[
\sigma\left(\omega\right)\sim\frac{\omega^{6}}{M^{8}}\,.
\]
In local thermal equilibrium at a temperature $T$, the number density of dynamical excitations
must be 
\begin{equation}
n\sim T^{3}\,,\label{eq:n-of-T}
\end{equation}
and the scattering rate is therefore 
\[
n\sigma\sim\frac{T^{9}}{M^{8}}\,.
\]
When this rate is comparable to the Hubble parameter, excitations
can no longer effectively thermalize, giving 
\begin{equation}
T_{\mathrm{fr}}\sim M\left(\frac{H}{M}\right)^{1/9}\,.\label{eq:T-fr}
\end{equation}
The freeze-out temperature gives, via Eq.\eqref{eq:n-of-T}, a minimum
$n$ required for thermalization. We can compute the number density
of \emph{dynamical} excitations directly from the $\left|\beta_{k}\right|^{2}$,
\begin{equation}
n\left(a\right)=\frac{1}{a^{3}}\int_{aH}^{aM}\left|\beta_{k}\right|^{2}d^{3}k\,.\label{eq:n-of-a}
\end{equation}
and when $n\left(a\right)\lesssim T_{\mathrm{fr}}^{3}$, new modes
can no longer thermalize. While dynamical excitations are constantly being added as modes enter
the EFT, they are also lost as modes cross the horizon, so that it is reasonable to suppose that $n\left(a\right)$ is
a decreasing function of the scale factor. Indeed, this is always true in a thermal phase. There then  exists a critical value
$a_{\mathrm{fr}}$ such that 
\begin{equation}
n\left(a_{\mathrm{fr}}\right)=T_{\mathrm{fr}}^{3}\,\label{eq:afr-def}
\end{equation}
and modes for which $k/M\lesssim a_{\mathrm{fr}}$ enter the EFT while
it is thermal, and equilibrate. For modes that are inside the horizon when thermal freeze-out occurs,  the $\left|\beta_{k}\right|^{2}$ satisfy 
\[
\left|\beta_{k}\right|^{2}=e^{-k/a_{\mathrm{fr}}T_{\mathrm{fr}}}\,,\,\frac{k}{M}<a_{\mathrm{fr}}\,.
\]
Modes for which $k/M>a_{\mathrm{fr}}$
enter the EFT \emph{after} freeze-out, and so their initial $\left|\beta_{k}\right|^{2}$
will not be significantly changed by interactions. They are therefore
of much greater interest, as they will be the ones carrying primordial information. 

\begin{figure}
\begin{centering}
\includegraphics[width=0.6\columnwidth]{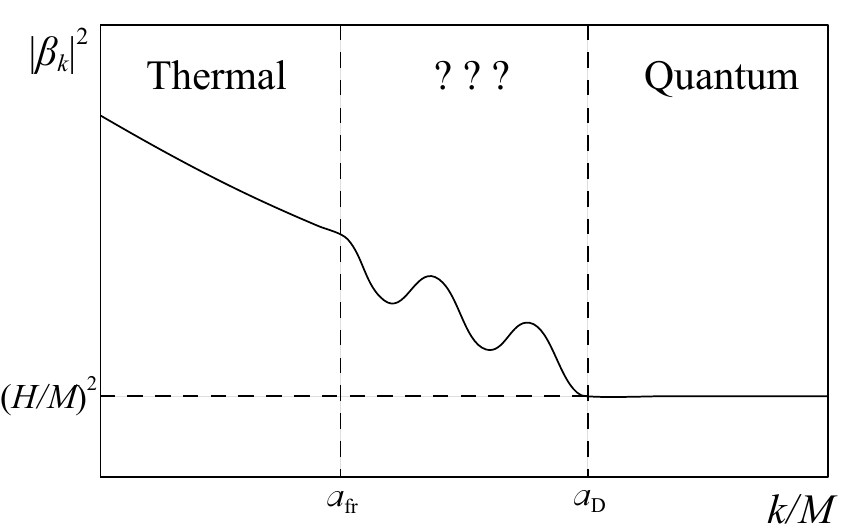}
\par\end{centering}

\caption{\label{fig:GeneralPlot}
A sketch of the late time $\left|\beta_{k}\right|^{2}$ when $a_\mathrm{fr}<a_\mathrm{D}$. Although the primordial $\left|\beta_{k}\right|^{2}$ are obscured for $k/M<a_\mathrm{fr}$ and $k/M>a_\mathrm{D}$, an observable window remains for $a_\mathrm{fr}<k/M<a_\mathrm{D}$.}
\end{figure}

The second obstacle to observing contributions to the $\left|\beta_{k}\right|^{2}$
from effects due to physics above the cutoff $M$ is that quantum effects also make a small contribution
to $\left|\beta_{k}\right|^{2}$. In \cite{Danielsson2002}, Danielsson argued
that this contribution is a constant, $\left(H/M\right)^{2}$. If the higher energy contributions
to the $\left|\beta_{k}\right|^{2}$ are smaller than\footnote{Other authors \cite{Bozza2003,Greene:2005wk} have argued for contributions smaller than $(H/M)^2$. Since we want to make a conservative estimate of the observability of initial conditions, we use the largest reasonable estimate for the quantum noise.} $\left(H/M\right)^{2}$,
they cannot be distinguished from quantum contributions and we regard
them as uninteresting. If we assume that the $\left|\beta_{k}\right|^{2}$
are decreasing functions of $k$, as would be required of a Hadamard state,  then there is a critical value of
the scale factor $a_{\mathrm{D}}$, 
\begin{equation}
\left|\beta_{k=a_{\mathrm{D}}M}\right|^{2}=\left(\frac{H}{M}\right)^{2},\label{eq:aD-def}
\end{equation}
beyond which quantum effects make the dominant contribution to the
$\left|\beta_{k}\right|^{2}$. So long as $a_{\mathrm{fr}}<a_{\mathrm{D}}$,
we find a window, 
\[
a_{\mathrm{fr}}<\frac{k}{M}<a_{\mathrm{D}}\,,
\]
where UV contributions to $\left|\beta_{k}\right|^{2}$ are distinguishable
from quantum effects, but do not lead to thermalization. This is illustrated
in Figure \ref{fig:GeneralPlot}. %

Simple considerations within the regime of validity of the EFT lead to our principal results: initial conditions for the inflaton cannot be specified on a spacelike hypersurface, the initial conditions may contain information about physics above the cutoff of the EFT, and in some circumstances this information is available at late times. Since we cannot construct a model of the $\left|\beta_k\right|^2$ using the EFT alone, all that we have assumed about the $\left|\beta_k\right|^2$ so far is that they are roughly decreasing. In order to illustrate the general arguments above, we now introduce a simple model for the $\left|\beta_k\right|^2$
whose effects remain observable at late times.


Without elaborating on UV completions that might give rise to these initial conditions, we consider a simple power law:
\[
\left|\beta_{k}\right|^{2}=\left(\frac{k}{m}\right)^{-\alpha}\,.
\]
Using \eqref{eq:n-of-a}, we find that the UV contribution to the
number density is 
\begin{eqnarray*}
n\left(a\right) & = & n_{0}a^{-\alpha}\,,\\
n_{0} & \sim & \begin{cases}
M^{3}\left(\frac{m}{M}\right)^{\alpha}, & 0<\alpha<3\,,\\
m^{3}\log\left(\frac{M}{H}\right), & \alpha=3\,,\\
H^{3}\left(\frac{m}{H}\right)^{\alpha}, & \alpha>3\,.\end{cases}
\end{eqnarray*}
We then use Eq.\eqref{eq:afr-def} and Eq.\eqref{eq:aD-def} to compute
\begin{eqnarray*}
a_{\mathrm{fr}} & = & \left(\frac{n_{0}}{M^{3}}\right)^{1/\alpha}\left(\frac{M}{H}\right)^{1/3\alpha},\\
a_{\mathrm{D}} & = & \frac{m}{M}\left(\frac{M}{H}\right)^{2/\alpha}.
\end{eqnarray*}
The UV contributions to $\left|\beta_{k}\right|^{2}$ are only observable
when $a_{\mathrm{fr}}<a_{\mathrm{D}}$, so we compute 
\[
\frac{a_{\mathrm{fr}}}{a_{\mathrm{D}}}\sim\begin{cases}
\left(\frac{H}{M}\right)^{5/3\alpha}, & 0<\alpha<3\,,\\
\left(\frac{H}{M}\right)^{5/9}\left[\log\left(\frac{M}{H}\right)\right]^{1/3}, & \alpha=3\,,\\
\left(\frac{H}{M}\right)^{-1+14/3\alpha}, & \alpha>3\,.\end{cases}
\]
Whenever $\alpha<14/3$, we find that $a_{\mathrm{fr}}<a_{\mathrm{D}}$
and so detailed observations of the initial conditions can, in principle,
be made. If $\alpha>14/3$, the late time $\left|\beta_{k}\right|^{2}$
interpolate from a thermal distribution to a flat $\left(H/M\right)^{2}$.

\section{Observational Consequences of Thermalization}
\label{EX}

We have argued that in the EFT of inflation a non-Bunch-Davies initial state is natural, and that it produces distinctive signatures at late times.  In this section we look more carefully at the prospects for observing those signatures.  A principal challenge is the diversity of possible initial states, in the absence of any \emph{a priori} constraints.  We will focus our attention on the consequences of an essentially universal phenomenon: thermal freeze-out. Assuming a fiducial value for $H/M=10^{-4}$, we demonstrate that the WMAP7 data can be used to constrain the other parameter in this simple model, the freeze-out momentum, and conclude that $k_\mathrm{fr}\gtrsim0.001\,\mathrm{Mpc}^{-1}$ is strongly disfavored.

The cosmic microwave background (CMB) provides two different ways to observe the $\left|\beta_{k}\right|^{2}$. 
Based on the findings of \cite{Chen2007,Holman2008}, we expect that non-vanishing $\left|\beta_{k}\right|^{2}$ lead to a distinctive form of non-Gaussianity, localized on ÒflattenedÓ triangles, i.e. those where $\left|\vec{k}_{1}\right|+\left|\vec{k}_{2}\right|\approx\left|\vec{k}_{3}\right|$. The amplitude of these contributions is proportional to $\left|\beta_{k}\right|$, and so the spectrum of flattened triangles provides a reasonably direct probe of the initial conditions. We believe this is the most direct way to reconstruct the initial conditions of the inflaton. Unfortunately, we currently have neither positive observations nor bounds on the existence of such triangles \cite{Komatsu2011,Meerburg:2009ys}, so we must look elsewhere to constrain the $\left|\beta_{k}\right|^{2}$.

As originally
noted by Danielsson \cite{Danielsson2002} (see also \cite{Easther2002}), non-standard initial conditions alter the scalar power
spectrum in a simple way:
\begin{equation}
P\to P\left(1+2\left|\beta_{k}\right|^{2}-2\mbox{Re}\left[\alpha_{k}^{*}\beta_{k}\right]\right).\label{eq:Danielsson}
\end{equation}
One might worry that current observations of the power spectrum stringently constrain the $\left|\beta_{k}\right|^{2}$.  Given a set of $\left|\beta_{k}\right|^{2}$, it is straightforward to compare with the available data, but as mentioned earlier we lack a well-motivated model. We therefore introduce a simplified model of the case where $a_\mathrm{fr}>a_\mathrm{D}$ (e.g. $\alpha>14/3$ in the power law example). As argued above, this does not permit a window into UV physics, but because of this it is much easier to use the power spectrum to constrain the model.

We will focus primarily on the thermal contribution to the $\beta_{k}$.
This universal contribution primarily affects low values of $k$,
which are the least sensitive to foreground
effects. We leave out the quantum effects
relevant for high $k$ as we expect them to be washed out by foreground
physics. Our model for the $\left|\beta_{k}\right|^{2}$ in the thermal
phase is then
\[
\left|\beta_{k}\right|^{2}=\exp\left[-\frac{k}{a_{\mathrm{fr}}T_{\mathrm{fr}}}\right].
\]
It is convenient to rewrite $a_{\mathrm{fr}}$ in terms of $k_{\mathrm{fr}}$,
the comoving momentum of the last mode to enter the EFT during the
thermal phase, 
\[
k_{\mathrm{fr}}=a_{\mathrm{fr}}M\,.
\]
Substituting in \eqref{eq:T-fr}, we then have 
\[
\left|\beta_{k}\right|^{2}=\exp\left[-\frac{k}{k_{\mathrm{fr}}}\left(\frac{H}{M}\right)^{-1/9}\right],\, k<k_{\mathrm{fr}}\,.
\]
We have given a model for $\left|\beta_{k}\right|^{2}$, but the last
term in \eqref{eq:Danielsson} is sensitive to the relative phase of $\alpha_{k}$ and $\beta_{k}$.
We assume that due to thermal effects the phases of different modes are uncorrelated,
and that the final term averages to zero. This leaves 
\[
P=P_{\mathrm{BD}}\times\begin{cases}
\left(1+2\exp\left[-\frac{k}{k_{\mathrm{fr}}}\left(\frac{H}{M}\right)^{-1/9}\right]\right) & k<k_{\mathrm{fr}}\,,\\
1 & k>k_{\mathrm{fr}}\,.\end{cases}
\]

We calculate $P_{\mathrm{BD}}$ using the CCFKS EFT. Retaining
only the leading term in the slow-roll expansion, 
\[
\epsilon=-\frac{d\log H}{d\log a}\,,
\]
we have 
\[
H=H_{0}\exp\left[-\epsilon\log\left(\frac{k}{k_{0}}\right)\right],
\]
and 
\[
P_{\mathrm{BD}}=A_{\mathrm{s}}\exp\left[\left(n_{\mathrm{s}}-1\right)\log\left(\frac{k}{k_{0}}\right)\right],
\]
with 
\[
n_{\mathrm{s}}-1=-2\epsilon\,.
\]
These are all standard results. They lead to our model power spectrum,
\begin{equation}
\log P\left(k\right)=\log A_{\mathrm{s}}+\left(n_{\mathrm{s}}-1\right)\log\left(\frac{k}{k_{0}}\right)+\begin{cases}
\log\left(1+2\exp\left[-\frac{k}{k_{\mathrm{fr}}}\left(\frac{H}{M}\right)^{-1/9}\right]\right) & k<k_{\mathrm{fr}}\,,\\
0 & k>k_{\mathrm{fr}}\,.\end{cases}\label{eq:Spectrum}
\end{equation}
Note that the last term is not naturally expanded in powers of $\log k$, and so the effects of thermal initial conditions cannot be recast as shifted values of $n_\mathrm{s}$ or $n_\mathrm{run}$.

\begin{figure}
\begin{centering}
\includegraphics[width=0.6\columnwidth]{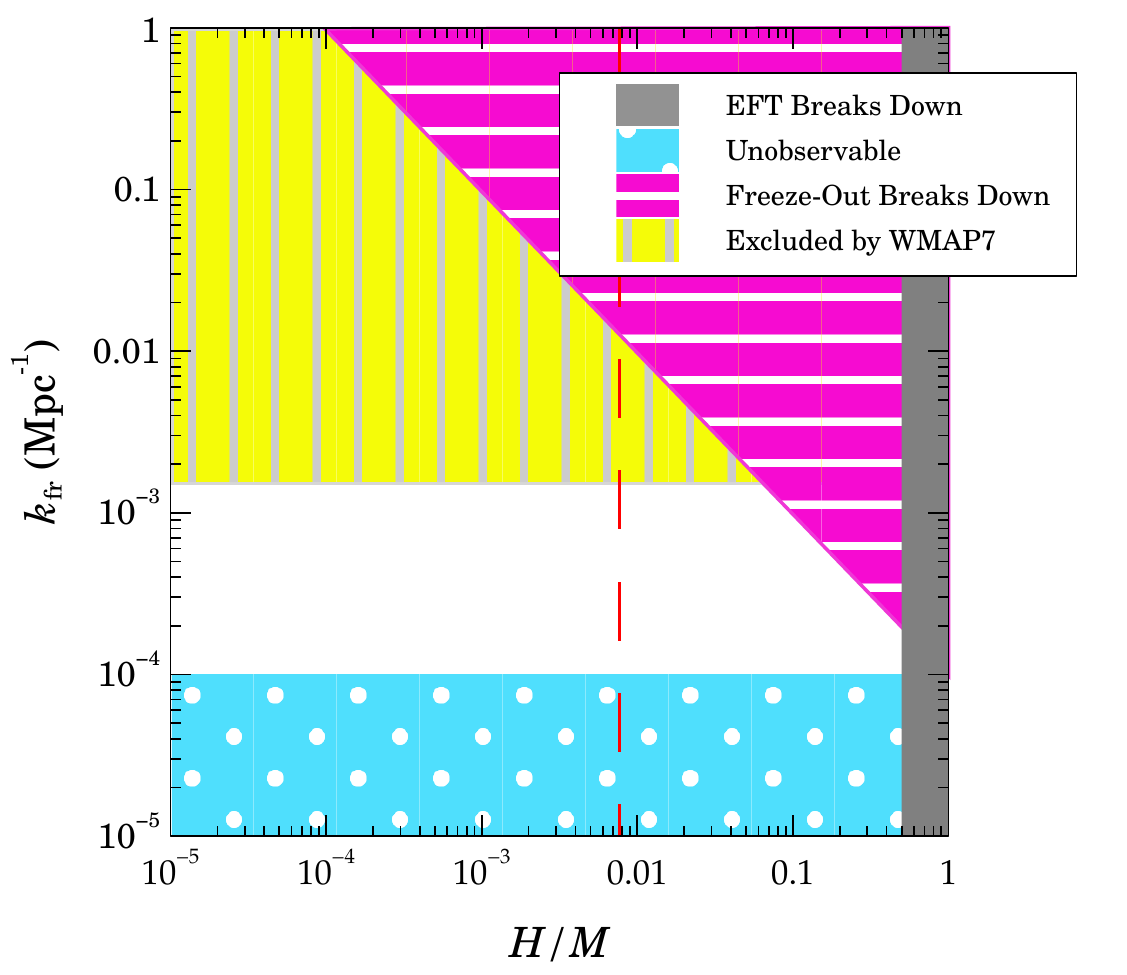}
\par\end{centering}

\caption{\label{fig:RangePlot}
Parameter space for the thermal freeze-out model. General considerations restrict the allowed values of $H/M$ and $k_\mathrm{fr}$, and comparison with the WMAP7 data restricts us to the window $10^{-4}\,\mathrm{Mpc}^{-1}\lesssim k_\mathrm{fr} < 1.5\times10^{-3}\,\mathrm{Mpc}^{-1}$. To the left of the red line the backreaction is automatically satisfied, while to the right it may or may not be, depending on the value of $H/M_\mathrm{Pl}$.}
\end{figure}

In addition to the usual parameters of the slow-roll power spectrum, the power spectrum in Eq.\eqref{eq:Spectrum} contains two additional parameters: $H/M$ and $k_\mathrm{fr}$.  General considerations restrict the allowed values of these parameters.  Since the ratio $H/M$ controls the range of validity of the EFT, we require that $H/M\lesssim0.5$.  In order for the effects of a a thermal phase to be observable, $1/k_\mathrm{fr}$ must be less than the size of the currently observable universe $R_\mathrm{obs}\sim10^4 \mathrm{Mpc}$, i.e. $k_\mathrm{fr}\gtrsim10^{-4}\mathrm{Mpc}^{-1}$. Finally, if the simple freeze-out model is to apply\footnote{In order to evolve back to earlier times, we would need to know the equation of state for the dynamical modes.  This is not just $w=1/3$, since modes become non-dynamical when they cross the horizon, carrying energy with them, and new modes constantly cross into the EFT. The latter effect is non-universal, since the energy density introduced by these modes is proportional to $\left|\beta_{k}\right|^{2}$.}, we must ensure that modes that crossed the horizon \emph{before} the proposed freeze-out are not part of the CMB sky.  Such modes are excluded if $a_\mathrm{fr} H<1/R_\mathrm{obs}$, which can be rewritten in terms of our model parameters as:
\[
k_\mathrm{fr}\frac{H}{M}\lesssim10^{-4}\,\mathrm{Mpc}^{-1}.
\]
The allowed range of values is indicated in Figure \ref{fig:RangePlot}.

In order to compare the freeze-out model with observation, we modified CAMB \cite{Lewis2000} to use
the power spectrum in Eq.\eqref{eq:Spectrum}. We then chose several test
values of $H/M$ and $k_{\mathrm{fr}}$, and used CosmoMC \cite{Lewis2002} to determine
best-fit values of the remaining parameters $A_{\mathrm{s}}$ and
$n_{\mathrm{s}}-1$, as well as $\Omega_{\mathrm{b}}h^{2}$, $\Omega_{\mathrm{DM}}h^{2}$,
$\theta,$ and $\tau$. 
Having determined best-fit values of the cosmological parameters for a given value of $k_\mathrm{fr}$, we compared the resulting power spectrum (generated by CAMB) with the binned WMAP7 observations \footnote{Available at \texttt{http://lambda.gsfc.nasa.gov/product/ map/dr4/pow\_tt\_spec\_get.cfm}}. We then assumed that the bins were statistically independent \footnote{This is not, strictly speaking, correct, but we don't believe it significantly affects the bounds on $k_\mathrm{fr}$.} and constructed a $\chi^2$ statistic. We plot the $3\sigma$-excluded region in Figure \ref{fig:RangePlot}.  The dependence on $H/M$ is quite weak, and it appears that freeze-out effects are only allowed and observable when 
\[
10^{-4}\,\mathrm{Mpc}^{-1}\lesssim k_\mathrm{fr} < 1.5\times10^{-3}\,\mathrm{Mpc}^{-1}\,.
\]
That is, the range of modes where the effects of thermal freeze-out are allowed is essentially the range of modes dominated by cosmic variance. 

We comment briefly on the issue of backreaction. Inflaton excitations could spoil the inflationary background if the associated energy density became comparable to $H^2 M_\mathrm{Pl}^2$. We rescale, and write this requirement as
\[
\left.H^{-4} \rho\right|_{a=a_{\mathrm{fr}}} \lesssim \frac{M_\mathrm{Pl}^2}{H^2}
\]
After integration, the lefthand side is a function of $H/M$ alone. Current observations of the scalar power spectrum and tensor-to-scalar ratio \cite{Larson2011} lead to the bound $H/M_\mathrm{Pl}\leq6.5\times10^{-5}$.  Unfortunately, this does not lead to a strict bound on $H/M$. We find that for $H/M\lesssim0.008$, the backreaction bound is automatically satisfied, while for  $H/M\gtrsim0.008$ it may or may not be, depending on the value of $H/M_\mathrm{Pl}$. 

Earlier in this paper we established two physical effects which might obscure signals from initial conditions. Our study of thermal freeze-out indicates that the WMAP7 observations are sufficient to exclude one of these, thermalization, from the CMB sky. Further observation, particularly of CMB non-Gaussianities, will let us look for signals that would contain more information about the UV theory.

\section{Conclusion}
\label{concl}

Our point in this letter was two-fold. First, we made the case that the standard flat space philosophy behind the use of effective field theory is insufficient in an expanding universe, or more generally, in any time dependent situation that allows a flow of modes across the UV cutoff of the EFT. This is reflected in the fact that an initial state must be specified for the fluctuations whose time dependent correlators we want to calculate. In flat space, where the relevant observables are scattering type amplitudes, it is sufficient to specify the asymptotic in/out vacuum states. In an FRW universe, however, the best we can do is specify the state of a given mode as it crosses {\em into} the domain of validity of the EFT, i.e. to scales below the UV cutoff $M$. 

Second, we showed that under certain reasonable assumptions, it may be possible to access {\em much} more information about the initial state in an inflationary cosmology than is possible in flat space. In particular, we argued that there was a window of opportunity between the thermalization of the information in the initial state and the swamping of this information due to quantum noise, where UV information could propagate to late time observables. Further, we have been able to use WMAP7 data to argue that the case where the window is closed due to thermalization of the fluctuations is already ruled out, making us optimistic that this window may be open rather than closed.

To tease the information about the inflationary initial state out of the CMB, more data are needed. Fortunately, the Planck mission will make many more modes of the CMB available to us, through better measurements of the power spectrum, non-Gaussianities, and some polarization information. These could all be used, perhaps in conjunction with 21 cm data, to allow us a glimpse into the deep UV sector of inflation.

\begin{acknowledgments}
It is a pleasure to acknowledge helpful discussions with Timothy Cohen and Matthew Johnson.  R.~H. was supported in part by the Department of Energy under grant DE-FG03-91-ER40682. 
\end{acknowledgments}

\bibliographystyle{JHEP}
\bibliography{EFT-Cosmo}

\end{document}